\documentclass[epj]{webofc}
\usepackage[varg]{txfonts}   
\wocname{EPJ Web of Conferences}
\woctitle{ICNFP 2016}
\newcommand{\bb}{$0\nu\beta\beta$}
\begin{document}
\selectlanguage{english}
\title{The CUORE and CUORE-0 experiments at LNGS}
%
%

\author{A.~D'Addabbo\inst{1}\fnsep\thanks{\email{antonio.daddabbo@lngs.infn.it}} \and C.~Alduino\inst{2} \and K.~Alfonso\inst{3}\and D.~R.~Artusa\inst{1,3}\and F.~T.~Avignone~III\inst{2}\and O.~Azzolini\inst{4}\and T.~I.~Banks\inst{5,6}\and G.~Bari\inst{7}\and J.W.~Beeman\inst{8}\and F.~Bellini\inst{9,10}
% ...
}

\institute{
INFN - Laboratori Nazionali del Gran Sasso, Assergi (L'Aquila) I-67010 - Italy \and
Department of Physics and Astronomy, University of California, Los Angeles, CA 90095 - USA \and
Department of Physics and Astronomy, University of South Carolina, Columbia, SC 29208 - USA \and
INFN - Laboratori Nazionali di Legnaro, Legnaro (Padova) I-35020 - Italy \and
Department of Physics, University of California, Berkeley, CA 94720 - USA \and
Nuclear Science Division, Lawrence Berkeley National Laboratory, Berkeley, CA 94720 - USA \and
INFN - Sezione di Bologna, Bologna I-40127 - Italy \and
Materials Science Division, Lawrence Berkeley National Laboratory, Berkeley, CA 94720 - USA \and
Dipartimento di Fisica, Sapienza Universit\`{a} di Roma, Roma I-00185 - Italy \and
INFN - Sezione di Roma, Roma I-00185 - Italy
% ...
}

\abstract{%
The Cryogenic Underground Observatory for Rare Events (CUORE) is a \mbox{1-ton} scale bolometric experiment devoted to the search of the neutrinoless double-beta decay (\bb) in $^{130}$Te. The CUORE detector consists of an array of 988 TeO$_2$ crystals operated at 10 mK. 
CUORE-0 is the CUORE demonstrator: it has been built to test the performance of the upcoming CUORE experiment and represents the largest $^{130}$Te bolometric setup ever operated. CUORE-0 has been running at Laboratori Nazionali del Gran Sasso (Italy) from 2013 to 2015. The final CUORE-0 analysis on $0\nu\beta\beta$ and the corresponding detector performance are presented. The present status of the CUORE experiment, now in its final construction and commissioning phase, are discussed. The results from assembly of the detector and the commissioning of the cryostat are reported.
}
\maketitle
\section{Introduction}
\label{intro}
Neutrinoless double-beta decay ($0\nu\beta\beta$) is a second order nuclear decay in which two beta decays occur simultaneously in a nucleus (even A, even Z), with the emission of two electrons and no neutrinos [$(A, Z) \rightarrow (A, Z+2) + 2e^-$]. This process was first hypothesized in~\cite{Goeppert-Mayer} and has never been observed so far. It would be possible only if the anti-neutrino ($\bar{\nu_e}$) emitted in one of the two beta decays, is re-absorber in the other one as a neutrino ($\nu_e$). The \bb~ therefore represents a window on some fundamental parameters in Neutrino Physics. Its observation would demonstrate that neutrinos are Majorana particles, that is neutrino and anti-neutrino are the same particle ($\nu_e = \bar{\nu_e}$), a feature that has never been so long osserved in any of the known matter constituent. Moreover, it would represent the first evidence of lepton number violation, supporting the leptogenesis as the matter-antimatter asymmetry generation mechanism. Finally, it would constrain the absolute neutrino mass scale, an information not accessible by oscillation experiments which can measure only the difference between the squared neutrino masses. By measuring the $0\nu\beta\beta$ decay rate $\Gamma_{0\nu}$, and so its half-life $T_{1/2}^{0\nu} = (\Gamma_{0\nu})^{-1}$, it is possible to measure the so-called ``effective Majorana mass'' $\langle m_{\beta\beta} \rangle = |\sum_{i = 1, 2, 3}{U_{ei}^{2} m_i}|$, where $m_i$ are the values of the three neutrino masses, and $U_{ei}$ is the first row of the leptonic mixing matrix between flavour and mass eigenstates. These two quantities are linked by the relation $\Gamma_{0\nu} \propto \frac{\langle m_{\beta\beta} \rangle^2}{m_e^2}$, where $m_e$ is the electron mass and the proportionality factor mainly depends on the nuclear matrix element adopted. Even without any $0\nu\beta\beta$ signal, it is possible to estimate a lower and an upper limit for $T_{1/2}^{0\nu}$ and $\langle m_{\beta\beta} \rangle$, respectively.

Given its importance, an intense experimental effort is ongoing to search for this decay in several nuclei~\cite{GERDA}. The Cryogenic Underground Observatory for Rare Events (CUORE)~\cite{CUORE}, presently in the final stages of construction at the Laboratori Nazionali del Gran Sasso(LNGS), will be one of the most sensitive upcoming $0\nu\beta\beta$ experiments. CUORE aims to reach a sensitivity on the half-life of  $0\nu\beta\beta$ of $^{130}$Te T$_{1/2}^{0\nu}$ $=$ 9.5$\cdot$10$^{25}$ yr in 5 years (at 90 \% C.L.). Section \ref{cuore} describes the CUORE experimental setup, including the cryostat and the final detector.

The first tower built following the CUORE procedures was operated at the LNGS as a stand alone experiment, named CUORE-0, from March 2013 to March 2015. CUORE-0 has been built on the experience gained from its demonstrator, CUORICINO, operated at the LNGS from 2003 to 2008. The CUORE-0 experiment was custom designed to validate the CUORE detector performances and to set a world best limit on the half life of $0\nu\beta\beta$ decay of $^{130}$Te. The CUORICINO and CUORE-0 techniques and results will be discussed in section \ref{cuoreprecursors}.

Thanks to the experience and lessons learnt by CUORE-0, CUORE is currently approaching the final stage of its assembly and commissioning. The latest achievements and upgrades towards the completion of the CUORE setup will be commented in section \ref{cuorecommissioning}.

\section{The CUORE experiment}
\label{cuore}

CUORE is a bolometric experiment devoted to the search of $0\nu\beta\beta$ in $^{130}$Te. It aims to monitor a total $^{130}$Te mass $M = 206 kg$, cooled down to 10 mK in an ultra low background environment.

\subsection{The detector}

The CUORE detector is an array of 988 TeO$_2$ bolometers arranged in a compact structure of 19 towers, each one containing 52 TeO$_2$ crystals, disposed on 13 floors. Each bolometer is a TeO$_2$ crystal with dimensions $5 \times 5 \times 5 ~cm^3$ and weight $750~g$. 

\begin{figure}[h]
\centering
\includegraphics[  width=0.9\linewidth,keepaspectratio]{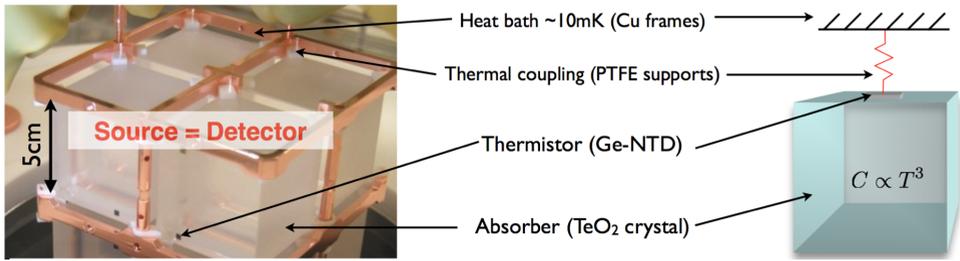}
\caption{A picture of a CUORE tower floor (4 TeO$_2$ crystals) with the corresponding bolometer scheme.}
\label{crystals} 
\end{figure}

Each bolometer consists of three main parts: an energy absorber, a temperature sensor, a weak thermal link ($G = pW/mK$) to the heat sink (see fig. \ref{crystals}). The energy provided by the decay ($\Delta E$) is released into the crystal, so that it acts either as the source and the absorber. This energy is then read by a Neutron-Transmutation Doped (NTD) Ge thermistor as a temperature increase $\Delta T \sim \Delta E/C$, where $C$ is the heat capacitance of the crystal. For $T=10mK \rightarrow C = 2 nJ/K$ and in turn the responsivity is $\sim 100 \mu K/MeV$. The thermistor temperature variation is converted into a variation of the electric resistance read as a voltage drop. This is amplified and readout at $125 Hz$ acquisition rate by a warm electronics. The energy release generates a pulse with one or more decay constants, with typical values of $C/G \sim 100 ms$. The pulse amplitude is calibrated using several emission peaks of a $^{232}$ Th source that can face the detector if necessary. The energy spectrum of the events is reconstructed using the different calibration peaks, whose principal is the 2615 keV line of $^{208}Tl$. The energy spectrum of the events we measure is actually the sum energy spectrum of the emitted electrons. As a consequence, the expected $0\nu\beta\beta$ signal is a peak centred at the Q-value of the $0\nu\beta\beta$ in $^{130}$Te, that corresponds to 2527.515 $\pm$ 0.013 keV \cite{Qval}. We usually refer to this portion of the spetrum as Region Of Interest (ROI).

\subsection{The cryostat}

The detector array is cooled down to 10 mK by a multistage cryostat unique of its kind: the first stage at $\sim 4K$ is obtained by means of 5 two-stage Cryomech PT415 Pulse Tubes (PTs), with nominal cooling power of 1.5 W at 4.2 K and 40 W at 45 K each.The base temperature is then provided by a DRS-CF3000 continuous-cycle $^3He/^4He$ Dilution Refrigerator (DR), customized by Leiden Cryogenics for CUORE purpouses. Its nominal cooling power is 5 $\mu$W at 12 mK (3 mW at 120 mK), and the minimum reachable temperature is around 5 mK (for more details on the CUORE cryostat see \cite{Ligi}). The dimensions ($> 10 ~m^3$), the mass ($\sim 25 tons$) and the cooling power ($3 \mu W$ at $10mK$) make it the today's biggest and most powerful dilution cryostat in the world.

\begin{figure}[h]
\begin{center}$
\begin{array}{c}
\includegraphics[%
  width=0.23\linewidth,keepaspectratio]{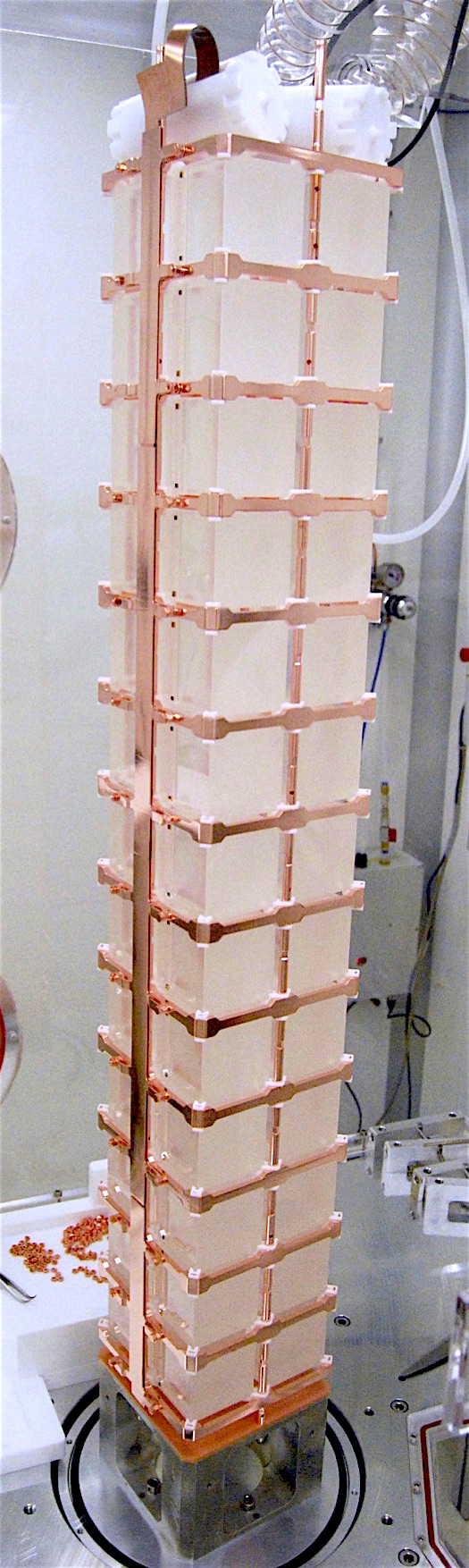}  
 \quad\quad
\includegraphics[%
  width=0.70\linewidth,keepaspectratio]{fig4.pdf}
\end{array}$
\end{center}
\caption{Left: the CUORE-0 detector during the assembling: the 52 TeO$_2$ crystals are arranged in 13 floors of 4 crystals each. Right: sketch of the CUORE cryostat with the CUORE detector surrounded by the vessels of increasing temperature and the two cryogenic lead shields.}
\label{C0picture}
\end{figure}

The cryostat (see Fig. \ref{C0picture}) is composed by 6 coaxial vessels, kept at 300 K, 40 K, 4 K, 600 mK, 50 mK, 10 mK, respectively. 300 K and 4 K shields are vacuum tight. The detector is placed inside the 10 mK vessel. The cryostat is suspended from above to a structure called Main Support Plate (MSP). A Y-Beam, isolated from the MSP through Minus-K  special suspensions, supports the detector. In this way, vibrations transmission from the cryostat to the crystals are minimized. To suppress external $\gamma$ background, two different lead shields are placed inside the cryostat (see sec. \ref{background}).



\subsection{The CUORE challenge}
\label{cuorechallenge}

The aim of future neutrinoless double-beta decay experiments, such as CUORE, is to approach the inverted hierarchy region of the neutrino masses. A very common way to evaluate the sensitivity of the measure is given by $T_{1/2}^{0\nu} \propto \sqrt{\frac{M \cdot t}{b \cdot \delta E}}$. Therefore, to achieve the necessary sensitivity, the key experimental parameters are: i) large mass of candidate nuclei $M$, ii) long live time $t$, iii) good detector energy resolution $\delta E$, and iv) very low radioactive background $b$. The CUORE goals for them are: M = 206 kg of $^{130}$Te, t = 5 years, $\delta E \leq 5 keV$ and $b \leq 10^{-2} c/keV/kg/y$.

The requirement i) is matched by the CUORE design. The status of detector assembling is reported in section \ref{cuorecommissioning}. 

To match challenges ii), iii) and iv) in CUORE, it is crucial the capability of cool down to 10 mK roughly 1 ton of detector, delivering a uniform and stable base temperature in a low vibration and low background environment. The CUORE cryostat has already proved its base temperature capability and stability over a long period. More details on the status of the cryostat commissioning can be found in section \ref{cuorecommissioning}. Thanks to the very low heat capacitance at low temperatures, the TeO$_2$ crystals has demonstrated a very good energy resolution ($\delta E/ E(ROI) \sim 0.002 \rightarrow \delta E \sim 5 keV$) at 10 mK (see section \ref{cuore0}). Here we briefly comment on the main background sources.

\subsection{Background} \label{background}

The excellent energy resolution of TeO$_2$ crystals makes the double beta decay with neutrinos emission ($2\nu\beta\beta$), that is the unavoidable intrinsic background source, easily distinguishable from $0\nu\beta\beta$ at the endpoint of the spectrum.

Besides that, one of the most common sources of background are the cosmic rays and the cosmogenic activation that they induce in the materials of the cryostat. For this reason the CUORE experiment is host in the Laboratori Nazionali del Gran Sasso (LNGS), an underground facility naturally shielded by roughly 1600 m of rocks ($\sim 3800~m.w.e.$) which provide a reduction of the atmospheric muons flux from $1.5 \times 10^{-2}~cm^{-2}s^{-1}$ to about $3 \times 10^{-8}~cm^{-2}s^{-1}$. 

Another relevant background source is given by the $\gamma s$ coming from the natural radioactive decay chain of $U$ and $Th$, that are present as contaminants in the cryostat. To suppress this contribution a set of lead shields has been installed inside the cryostat. Suspended between the 10mK plate and the TSP, it is placed a 30-cm-thick cylindrical shielding, made of lead sheets. This is the only shielding wall from the top side for a total weight of about 2 tons. It is hang from the 10 mK by the mean of three kevlar strips and it is thermalized at 50 mK. Another cold lead shielding, kept at 4 K, is a 6-cm-thick wall that surrounds the 600 mK vessel. The lateral shield is made of about 5 tons of $^{210}$Pb-free ancient roman lead. The aim of the cold shields is to prevent radioactivity contamination from the external vessels (300, 40, and 4 K) and from all the cryostat plates. This contribution is suppressed by accurate radio-purity controls on the materials, with a background evaluated in $\sim 6 \times 10^{-3} c/keV/kg/y$ by the CUORE-0 experiment (see sec. \ref{cuore0}). To shield the detector for the $\gamma s$ sources external to the cryostat, an external lead shield ($\sim$ 80 tons, thickness = 25 cm) can be raised in operation to dress the outer cryostat vessel at room temperature.

The main and more critical contribution in the ROI is due to the so-called ``degraded alphas''. These are $\alpha$-particles emitted by surface contaminations of the crystals and of the copper of which the tower structure is made. These $\alpha s$ release part of their energy in the surface that is emitting them (source), and the rest in the one that absorbs them (target). We refer to such an process as to a ``source-target'' event. Through anti-coincidence analysis is it possible to reject the ``crystal-crystal'' events, where both the source and the target are active surfaces, but it is not possible to tag ``crystal-copper'' nor ``copper-crystal'' events, which give rise to a continuos background in the ROI. 

The reduction of this contribution was a crucial part of the intense CUORE R\&D program, which is discussed in section \ref{cuoreprecursors}. The main improvements have been: i) the increased radio-purity of the copper and crystal surfaces as well as in the assembly environment, ii)  thicker shields using low-activity ancient Roman lead, and iii) the granularity and self shielding of the 19-towers array affords providing a better anti-coincidence coverage (a discussion of the different contribution can be found in \cite{ReD}). Thanks to these improvement, the background level in CUORE is expected to be as good as 0.01 counts/keV/kg/y \cite{Cbackground}.

\section{The CUORE precursors}
\label{cuoreprecursors}

CUORE has been designed on the experience of the predecessor experiments CUORICINO~\cite{QINOdet} and CUORE-0 ~\cite{C0det}, operated at the LNGS from 2003 to 2008, and from 2013 to 2015, respectively. 

\subsection{The CUORICINO experience}
\label{cuoricino}

The demonstrator experiment CUORICINO \cite{QINOdet} 
was composed by 62 TeO$_2$ bolometers, for a total mass of 40.7 kg and an acquired statistics of 19.75 kg($^{130}$Te) $\cdot$ y. The average energy resolution of the detector at 2615 keV ($^{208}$Tl line) was 6.3 $\pm$ 2.5 keV~\cite{QINOres} and the background level in the $0\nu\beta\beta$ energy region was 0.169 $\pm$ 0.006 c/keV/kg/y. No $0\nu\beta\beta$ signal was found. The corresponding lower limit on the $0\nu\beta\beta$ half-life of $^{130}$Te is 2.8 $\times$ 10$^{24}$ y (90\% C.L.). This limit translates into an upper limit on the neutrino effective Majorana mass ranging from 300 to 710 meV, depending on the nuclear matrix elements considered in the computation.

The scaling from CUORICINO to CUORE aims at improving the sensitivity to the $0\nu\beta\beta$ half life of $^{130}$Te. This goal can be achieved by increasing the exposure (increasing the active mass) and by reducing the background in the ROI. After CUORICINO, the achievement of these goals were part of the intense CUORE R\&D program.

A challenge related to the detector performances is the uniformity of their behaviour. A crucial requirement is the capability of completing 988 TeO$_2$ bolometers and assembly them into 19 towers in clean and reproducible conditions. In CUORICINO, we observed a large spread in the pulse shape among the bolometers. This leads to a complication in the data analysis that could be more challenging in case of an experiment, like CUORE, with about 1000 single detectors. 

For these reasons, before starting the construction of the 19 CUORE towers, an additional tower, named CUORE-0, was produced according to the CUORE requirements.

\subsection{The CUORE-0 experiment}
\label{cuore0}

CUORE-0 is a single CUORE-like tower, the first one built using the low-background assembly techniques developed for CUORE~\cite{Buccheri}. It is made of 52 TeO$_2$ bolometers, for a total mass of 39~kg. To improve the detector response uniformity special care has been devoted in the realisation of a new detector assembly system. 

Each TeO$_2$ detector is instrumented with a NTD-Ge thermistor glued on its surface, to measure the temperature change of the absorber and convert it into an electric signal.  The thermal response of the detector depends on its temperature. For this reason each crystal is instrumented also with a silicon resistor, called ``heater'', to generate reference pulses of $\sim$ 3 MeV. 
The tower was operated in Hall A of LNGS, in the same dilution refrigerator that previously hosted the CUORICINO experiment, and it ran between March 2013 and March 2015. Technical details are reported in~\cite{InitialPerformance} and~\cite{C0det}, while the CUORE-0 physics results can be found in~\cite{C0result}.

\subsubsection{Detector performance}

The detector performances of the CUORE-0 bolometers have been improved with respect to CUORICINO (see table \ref{tab-1}). One of the main improvement is a new layout of the detector towers, where the 52 TeO$_2$ crystals are held in an ultra-pure copper frame by PTFE supports and arranged in 13 floors, with 4 crystals per floor (see fig.~\ref{C0picture}). In order to reduce to the background events in the ROI, more stringent material selection criteria and strictly cleaning and handling procedures were implemented for the detectors and for the cryostat components, as well as for the assembly environment. 
Moreover, dedicated processes have been developed to suppress the degraded-$\alpha s$ induced background, increasing the radio-purity of the copper and crystal surfaces. Finally, thicker shields using low-activity ancient Roman lead have been used for the first time in the CUORE-0 cryostat.

CUORE-0 acquired data for the $0\nu\beta\beta$search collecting a total exposure of 9.8 kg$\cdot$y of $^{130}$Te. Data are grouped in month-long blocks called datasets. At the beginning and end of each dataset we calibrate the detector by placing a $^{232}$Th source next to the outer vessel of the cryogenic system.
 
We use the calibration line with the highest intensity and next to the ROI, 2615~keV from $^{208}$Tl, in order to study the detector response function to a mono energetic energy deposit for each bolometer and dataset. We estimate the shape parameters of the 2615~keV line with a simultaneous, Unbinned Extended Maximum Likelihood (UEML) fit to calibration data. The physics exposure-weighted effective mean of the FWHM values for each bolometer and dataset is 4.9 keV, with a corresponding RMS of 2.9 keV. We evaluate the background level in the alpha-dominated region (2700-3900)~keV to be 0.016 $\pm$ 0.001 counts/keV/kg/y), 6 times smaller with respect to the CUORICINO background in the same region (see fig.~\ref{C0res} left). 

\begin{table}[h]
\quad\quad\quad\quad\quad
\centering
\caption{A comparison between the CUORICINO and CUORE-0 resolution and background performances.}
\label{tab-1}       
\begin{tabular}{llll}
\hline
 & Resolution [keV]  & \multicolumn{2}{c}{Background [c/keV/kg/y]}  \\
 & at $^{208}$Tl peak (2615 keV) & 2.7-3.9 MeV  & ROI  \\\hline
CUORICINO & 6.3~~~~FWHM rms = 2.5 & 0.110 $\pm$ 0.001 & 0.169 $\pm$ 0.006 \\
CUORE-0 & 4.9~~~~FWHM rms = 2.9  & 0.016 $\pm$ 0.001 & 0.058 $\pm$ 0.004 \\\hline
\end{tabular}
\quad
\end{table}

\newpage
\subsubsection{$0\nu\beta\beta$ decay result}
\begin{figure}[h]
\begin{center}$
\begin{array}{c}
\includegraphics[%
  width=0.51\linewidth,keepaspectratio]{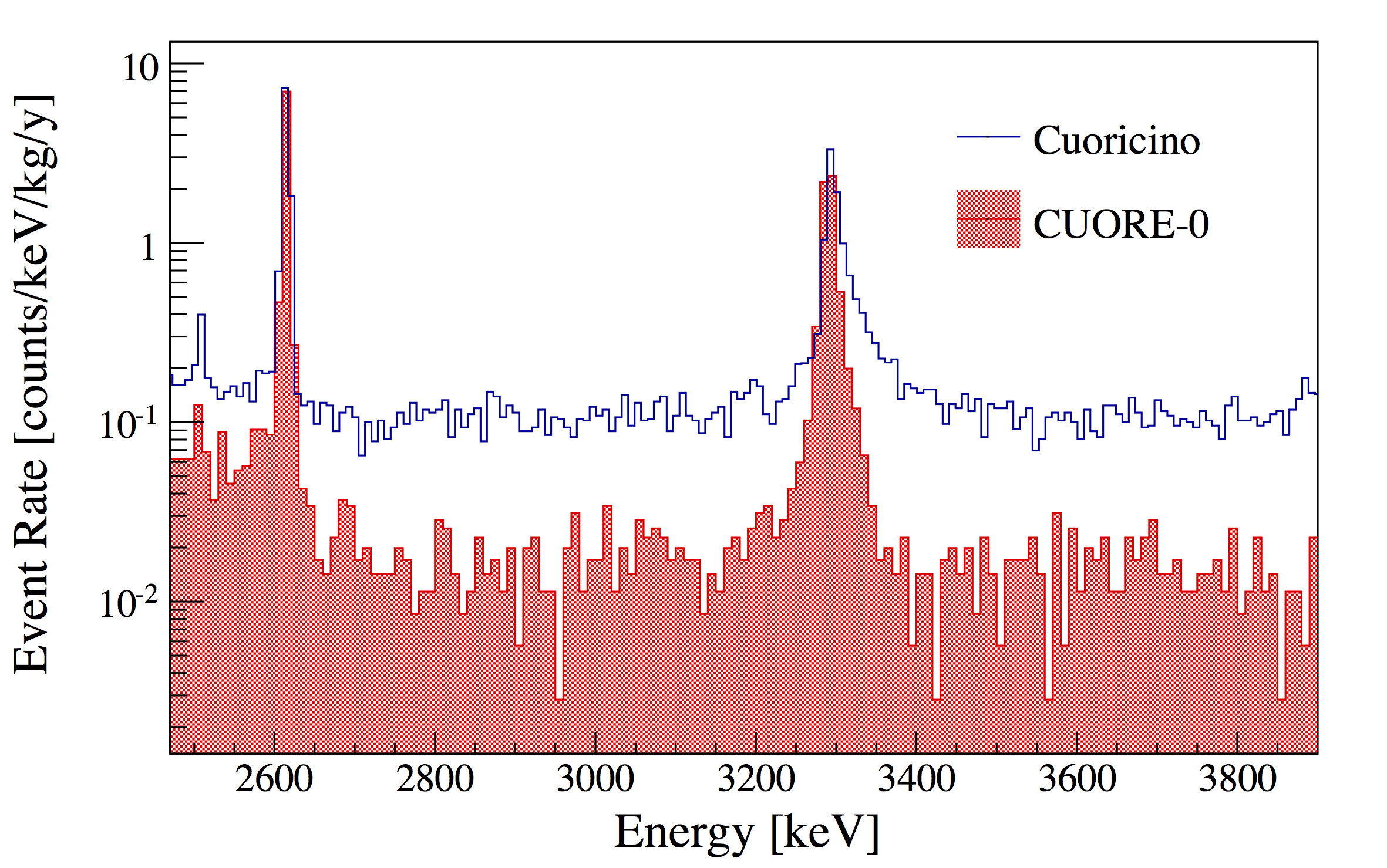}  
\includegraphics[%
  width=0.49\linewidth,keepaspectratio]{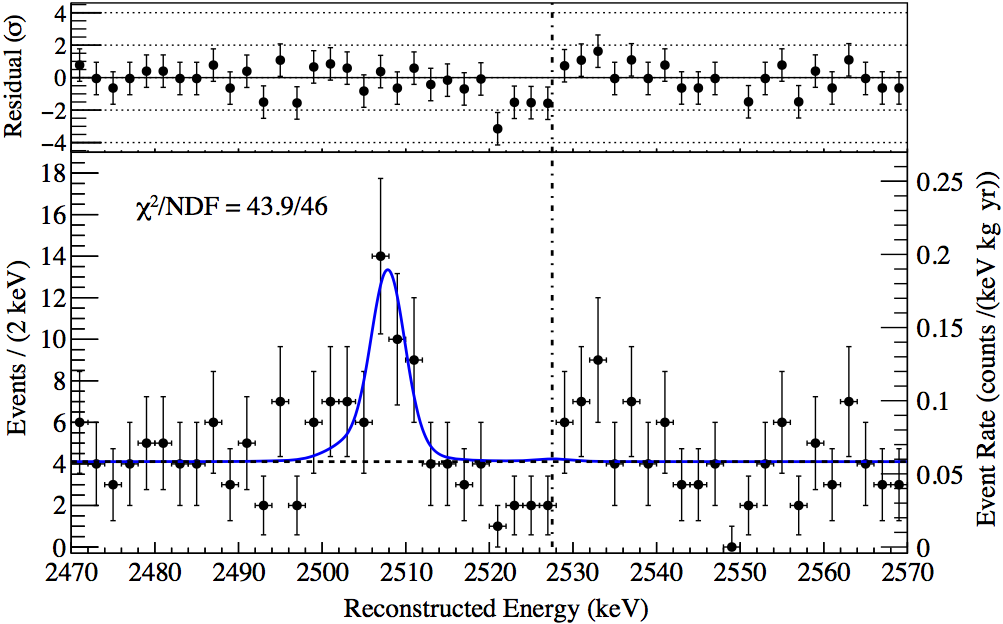}
\end{array}$
\end{center}
\caption{Left: Comparison of CUORE-0 and CUORICINO spectra in the flat alpha region (2700-3900 keV). Right bottom: The best-fit model from the UEML fit (solid blue line) overlaid on the spectrum of $\beta\beta$0$\nu$ decay candidates in CUORE-0 (data points); the data are shown with Gaussian error bars. Dotted black line shows the continuum background component of the best-fit model. Right top: normalized residuals of the best-fit model and the binned data. The vertical dot-dashed black line indicates the position of Q$_{\beta\beta}$}
\label{C0res}
\end{figure}

We search for $0\nu\beta\beta$ signal in $^{130}$Te in the final CUORE-0 energy spectrum performing a simultaneous UEML fit in the energy region 2470-2570~keV (Fig.~\ref{C0res} right). The fit function is composed by three parameters: a posited signal peak at the Q-value of the transition, a peak at $\sim$ 2507~keV from $^{60}$Co double-gammas, and a smooth continuum background attributed to multi-scatter Compton events from $^{208}$Tl and surface decays.
The best-fit values are $\Gamma_{0\nu}$ = 0.01 $\pm$ 0.12(stat) $\pm$ 0.01(syst) $\times$ 10$^{-24}$yr$^{-1}$ for the 0$\nu\beta\beta$ decay rate and 0.058 $\pm$ 0.004(stat) $\pm$ 0.002(syst) counts/keV/kg/y for the background index in the ROI. This result is 3 times lower than the Cuoricino background, 0.169 $\pm$ 0.006 counts/keV/kg/y, in the same ROI. 
Using a Bayesian approach, we set a 90\% C.L. lower bound on the decay half-life of 2.7 $\times$ 10$^{24}$yr~\cite{C0result}.
When combined with the 19.75 kg$\cdot$y exposure of $^{130}$Te from the Cuoricino experiment, we find a Bayesian 90\% C.L. limit of T$_{0\nu}$$>$ 4.0$\times$10$^{24}$yr, which is the most stringent limit to date on the $^{130}$Te 0$\nu\beta\beta$ half-life. Additional details on the analysis techniques can be found in~\cite{C0halflife}.

\section{CUORE commissioning}
\label{cuorecommissioning}

The commissioning of the CUORE experiment has two main deliverables: i) the cryostat capability to cool the detector down to the expected base temperature ($<$10 mK), in stable conditions and for a long period, and ii) the assembly and installation of the 988 TeO$_2$ bolometric detectors at the coldest stage of the cryostat.

The commissioning of the CUORE cryostat was concluded in March 2016. It consisted in a sequence of could run of increasing complexity in the cryostat. The last cold run was performed with the full cryostat with all the components installed but the detector. An 8 TeO$_2$ bolometer array, named Mini-Tower, was built, mounted, and connected to the electronics to check bolometer functionality. The CUORE cryogenic system demonstrated to be able to cool down an experimental volume of $\sim 1m^3$ at a base temperature of $6.3 mK$ for more then 70 days, with temperature variations of the order of $0.2 mK$ rms, a really important and unprecedented experimental achievement. The performance of the detectors measured on the Mini-Tower were encouraging  and a full debugging of the electronics, DAQ, temperature stabilization, and detector calibration systems was performed. 

The assembly of the CUORE detector towers started in late February 2013 and has been completed in July 2014. Following the experience of CUORE-0, the assembling was performed under nitrogen atmosfere in a dedicated assembling line. Each tower has been stored inside an individual cage and constantly flushed with nitrogen until their installation. The installation of the detectors in the CUORE cryostat started at the beginning of August 2016 and has been successfully completed on 26th August. Strictly cleaning procedures and a radon abatement system were adopted during all the installation period.

We are currently closing the cryostat vessels and finalizing all the subsystems. The final CUORE cool down will start by the end of November 2016.

\begin{figure}[h]
\begin{center}$
\begin{array}{c}  
\includegraphics[%
  width=0.47\linewidth,keepaspectratio]{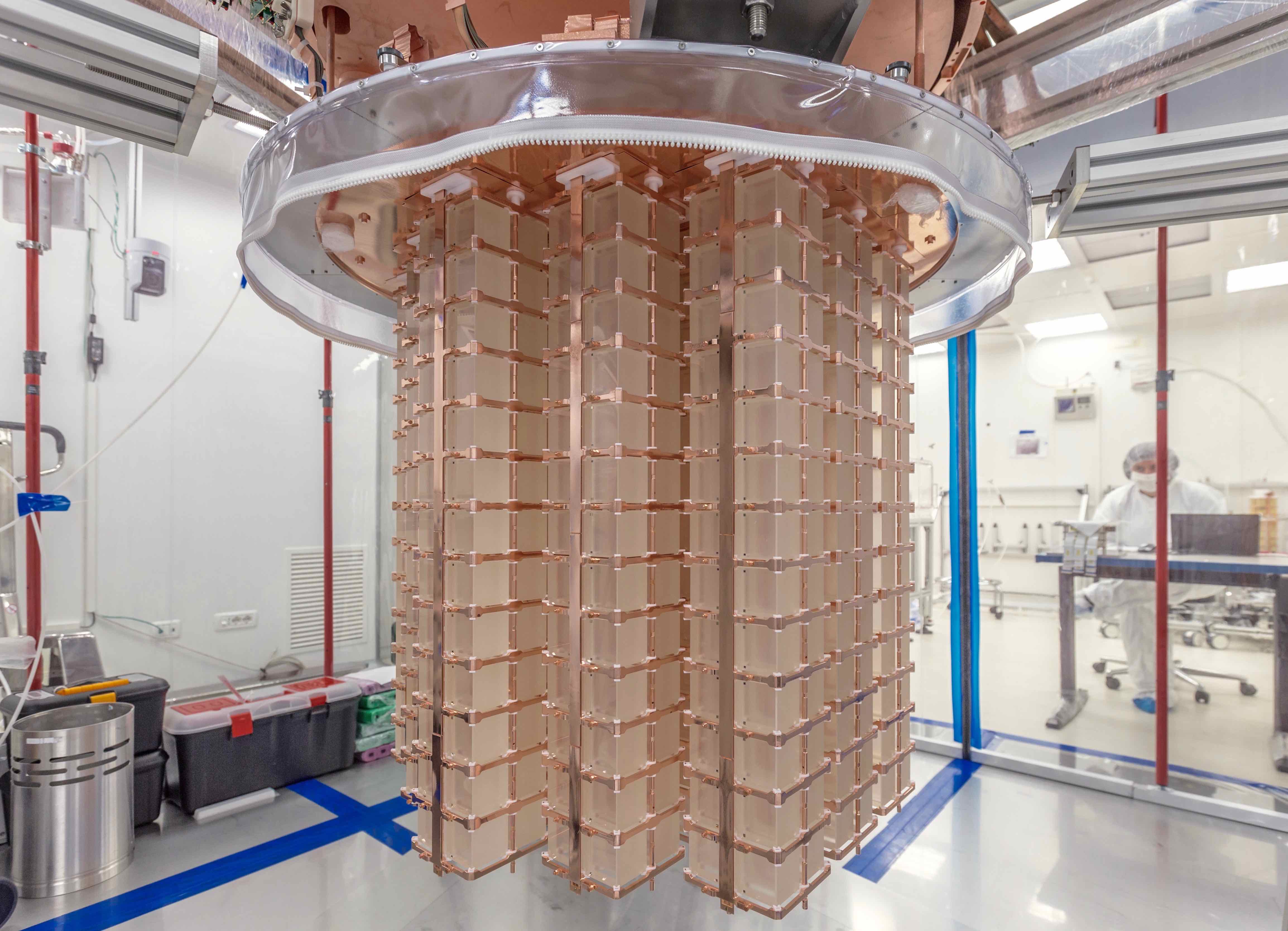}
~~
 \includegraphics[%
  width=0.52\linewidth,keepaspectratio]{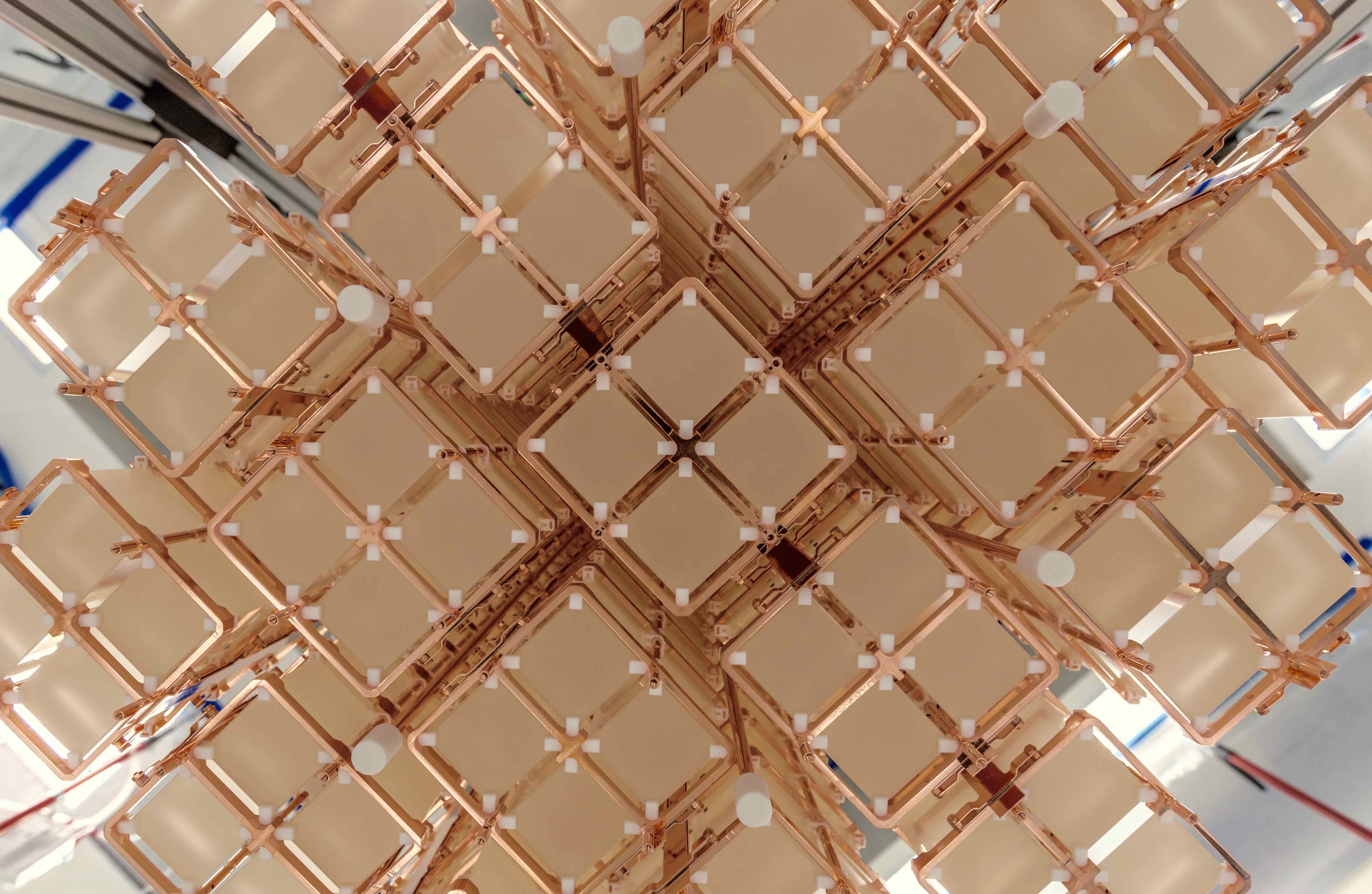}
\end{array}$
\end{center}
\caption{Two pictures of the CUORE detector after its installation in August 2016. Left: a picture of the detector attached to the MSP inside the CUORE cryostat cleanroom. Right: a bottom view of the 19 CUORE towers.}
\label{detector}
\end{figure}




%

\section*{Acknowledgments}

The CUORE Collaboration thanks the directors and staff of the Laboratori Nazionali del Gran Sasso and the technical staff of our laboratories. This work was supported by the Istituto Nazionale di Fisica Nucleare (INFN), the National Science Foundation, the Alfred P. Sloan Foundation, the University of Wisconsin Foundation, and Yale University. This material is also based upon work supported by the US Department of Energy (DOE) Office of Science and by the DOE Office of Nuclear Physics. This research used resources of the National Energy Research Scientific Computing Center (NERSC). More details can be found at: http: http://cuore.lngs.infn.it/support

%
%




\end{document}